\documentstyle[prl,aps]{revtex}
\begin{document}


\title{Linear vs. nonlinear coupling effects in single- and multi-phonon atom-surface scattering}
\author{A. \v{S}iber and B. Gumhalter}
\address{Institute of Physics of the University, P.O. Box 304,
10001 Zagreb, Croatia. }
\date{\today}
\maketitle

\begin{abstract}
We present a comparative assessment of the features of inelastic atom-surface scattering spectra that are produced by several different forms of linear and nonlinear phonon coupling to the projectile atom. 
Starting from a simple theoretical model of atom-surface scattering and employing several recently developed exact numerical and approximate analytical methods we calculate and compare the scattering probabilities ensuing from each form of interaction and from each calculational scheme. This enables us to demonstrate that in the regime of thermal energy atom scattering from surfaces the dominant contributions to the zero-, one- and multi-phonon excitation probabilities obeying unitarity arise from linear coupling treated to all orders in the interaction. 
\end{abstract}

\begin{center}
{\bf Published in Phys. Rev. Lett. 90, 126103 (2003)}
\end{center}


\newpage

\newcommand{\ga}{\: \raisebox{-.8ex}{$\stackrel{\textstyle
>}{\sim }$}\:\: }


Theoretical interpretations of inelastic atom-surface scattering experiments carried out in the past two decades to investigate vibrational properties of the various surfaces\cite{revs} have all been based on several assumptions concerning atom-surface interactions that have never been critically examined or discussed. 
The aim of the present Letter is to remedy this situation in one of its so far neglected aspects, viz. the lack of a detailed assessment of the effects that arise from different forms of atom coupling to quantized surface vibrations or phonons. 
We perform this task using a simple paradigmatic, yet realistic enough quantum model comprising all the essentials of single- and multi-phonon atom-surface scattering. 
  
The first assumption adopted by the majority of the authors in the development of inelastic atom-surface scattering theories reviewed in Refs. \onlinecite{plrep,JPhys,BenValb,BortoLevi,Hulpke,Doak,Santoro} has been the linear coupling of scattered atoms to vibrational displacements of surface atoms (i.e. linear coupling to the phonon field), and only few authors have resorted to descriptions based on nonlinear coupling\cite{KasaiBrenig,MansonJPhys}. 
The absence of a quantification of the effects originating from the different forms of coupling is rather surprising because they should manifest themselves in {\em all} the components of the scattering spectra, viz. in {\em (a)} elastic scattering probability given by the so-called Debye-Waller factor (DWF), {\em(b)} one-phonon scattering probability, and {\em (c)} multi-phonon scattering probabilities. 

The second assumption concerns the treatment of one-phonon excitation probabilities. Here, a rather general consensus prevailed that these should be calculated in the limit of linear atom-phonon coupling and in the Born approximation using distorted waves accounting for reflections of the atom from the surface to all orders in the static atom-surface potential\cite{BortoLevi,Santoro}. 
However, as the distorted wave Born approximation (DWBA) yields non-unitary scattering amplitudes it can provide reliable estimates only for the relative magnitudes of single quantum excitation probabilities in the regime in which the multi-quantum excitations are negligible. In order to obtain  accurate one-phonon excitation probabilities for comparison with  the zero- and multi-quantum ones (which are all sensitive to the form of atom-phonon coupling)  and with the experimental data, the DWBA results should be unitarised to comply with the optical theorem. 

The most striking differences arising from the assumptions of either linear or nonlinear atom-phonon coupling are expected in the calculated multiphonon scattering probabilities. Recent He atom scattering (HAS) experiments have successfully revealed and disentangled the one- and multi-quantum features in the atom-surface scattering spectra for phonons with Einstein\cite{XeCu,multiph} and acoustic\cite{Xe111,zone} type of dispersion. 
Hence, the resolution of the origin of multiquantum features 
in the measured spectra as being due either to {\em (i)} multiple single phonon scattering events describable in terms of linear coupling, or {\em (ii)} multiquantum excitations that arise from nonlinear coupling in each order of perturbation expansion, emerges as a task of utmost interpretative importance.

One of the important conditions for a conclusive testing of the roles that linear and nonlinear coupling may play in the various theoretical descriptions of atom scattering from surface phonons is that they all ought to be applicable to a model which realistically describes the essentials of a particular scattering system. Another condition is that a clear disentanglement of the single- and multi-quantum features is possible for comparison of the results of different theories with the available experimental evidence. 
Experimental systems that enable 
the easiest identification of and distinction between the various one- and multi-phonon excitations in atom-surface collisions are those in which atoms are inelastically scattered by dispersionless or Einstein phonons. 
In this case the inelastic scattering spectra display a series of well separated equidistant peaks at the multiples of Einstein phonon energy $\pm\hbar\omega_{0}$ away from the elastic line. 
The weight of the $n$-th peak measures the probability of excitation of $n$ phonons in the course of scattering.
A prototype collision system of this kind is provided by HAS from monolayers of Xe atoms adsorbed on Cu(111) and Cu(001) surfaces which support vertically polarized Einstein modes\cite{XeCu}. Other systems with similar properties have also been reported in the literature\cite{MansonJPhys,multiph}. 
The model Hamiltonians required for a full quantum mechanical description
of these systems have been presented
elsewhere\cite{plrep,XeCu}. However, for the present purpose all the relevant features of collision dynamics are included in the effective one-dimensional Hamiltonian $H$ that describes motion of a projectile nonlinearly coupled to a surface oscillator of frequency $\omega_{0}$. 
Denoting by $z$ $(Z)$, $p$ $(P)$ and $m$ $(M)$ the coordinate, momentum and mass of the scattered particle (surface oscillator), respectively, we can write 

\begin{eqnarray}
H&=&\frac{p^{2}}{2m}+\frac{P^{2}}{2M}+\frac{M\omega_{0}^{2}Z^{2}}{2}+ De^{-\alpha(z-Z)}\nonumber\\
&=& \frac{p^{2}}{2m}+ De^{-\alpha z}+\frac{P^{2}}{2M} +\frac{M\omega_{0}^{2}Z^{2}}{2} +De^{-\alpha z}(e^{\alpha Z}-1)\nonumber\\
&=&
H_{0}^{p}(p,z)+H_{0}^{osc}(P,Z)+V(z,Z).  
\label{eq:H}
\end{eqnarray}
Here, the elastic particle motion in the model static surface potential  of the Born-Mayer form, $U(z)=De^{-\alpha z}$, is described by distorted waves\cite{JM} that are eigenstates of the particle Hamiltonian $H_{0}^{p}(p,z)=p^{2}/2m+U(z)$. By introducing a purely repulsive $U(z)$ we avoid the occurrence of resonant scattering effects associated with the bound states of the surface potential well\cite{recovery,commAnd,kinfoc} that are unimportant for the present study.  
 The inelastic scattering is governed by the interaction potential 

\begin{eqnarray}
V(z,Z)&=&De^{-\alpha z}(e^{\alpha Z}-1)\nonumber\\
&=&
Z\alpha D e^{-\alpha z}+ \frac{Z^{2}\alpha^{2}}{2} D e^{-\alpha z}+ \dots \nonumber\\
&=&
ZV_{1}(z)+Z^{2}V_{2}(z)+\dots ,
\label{eq:Vexpanded}
\end{eqnarray}
that contains powers of oscillator displacements $Z$ to all orders.
The main merit of the thus defined model lies in its  amenability to exact numerical\cite{KasaiBrenig,CC} and approximate analytical treatments\cite{plrep} in the calculations of single- and multi-phonon scattering probabilities. This enables us to  consistently investigate as how the different forms of atom-phonon coupling affect the scattering spectra.

In this Letter the exact numerical treatments of elastic and inelastic atom scattering probabilities are carried out by employing the coupled channels  method\cite{CC}(CC) to solve the Schr\"{o}dinger equation of particle motion perturbed by three types of interactions embodied in expression (\ref{eq:Vexpanded}): the linear coupling term $Z V_{1}(z)$, the sum of linear and quadratic coupling terms $Z V_{1}(z)+Z^{2}V_{2}(z)$, and the full potential $V(z,Z)$. 
Tractable analytical solutions for the same model are obtained for the cases of linear coupling $Z V_{1}(z)$ treated to {\em all orders} within the distorted wave exponentiated Born approximation (EBA) formalism\cite{plrep} based on cumulant expansion, and for the full coupling $V(z,Z)$ treated in the DWBA as regards the particle motion. 
With the exception of the latter approximation, all other mentioned treatments yield unitary results in that the sum of all calculated scattering probabilities  equals unity.  Separate treatment of the sum of linear and quadratic coupling terms is particularly instructive because the quadratic term gives 
rise to both inelastic and elastic scattering processes, viz. the simultaneous two phonon excitations and elastic scattering by the static potential $V_{2}(z)$ renormalised by emission and reabsorption of a virtual phonon, an effect that is not present in linear coupling theories. These features are diagrammatically illustrated in Fig. \ref{nonlinscfg1} in which the virtual phonon excitation is represented by a closed phonon loop.
Diagrammatic perturbation expansion in $V(z,Z)$ shows that nonlinear coupling to all orders gives rise to the appearance of arbitrary number of closed phonon loops in any single- or multi-phonon vertex\cite{dwf1}. This leads to a Holstein type of renormalisation\cite{Holstein} of inelastic particle scattering matrix elements by the factor $\exp[-(\alpha u_{0})^{2}/2]$ where $u_{0}^{2}$ is the mean square displacement of the oscillator.  

The calculations have been carried out for the elastic scattering probability $P_{0}$, the one-phonon scattering probability $P_{1}$, and the two-phonon scattering probability $P_{2}$ for the types of couplings and within the  theoretical approaches described above by using the parameters representative of He and Ne atom scattering from adsorbed Xe atoms. 
Figure \ref{nonlinscfg2} compares the behaviour of $P_{n}$ $(n=0,1,2)$ calculated for linear, linear and quadratic, and full nonlinear couplings as the functions of scattered particle incoming energy $E_{i}$, for He atom scattering from Xe atoms adsorbed on a cold Cu surface. 
The generalisation to finite surface temperatures is straightforward\cite{KasaiBrenig,recovery} but does not bring in new effects concerning the various types of atom-phonon coupling. 

The important message that can be deduced from the behaviour of elastic scattering probability $P_{0}$ or the Debye-Waller factor depicted in Fig. \ref{nonlinscfg2} is an excellent agreement between the results of exact CC calculations for the sum of linear and quadratic and full nonlinear coupling in the whole range of the projectile incoming energy. This means that the nonlinear coupling effects are described to a high degree of accuracy already by the sum of linear and quadratic coupling terms. 
The EBA formalism\cite{plrep} that is based on approximate treatment of linear coupling to all orders is also in excellent agreement with the exact numerical results. 

Almost identical conclusions pertain to the behaviour of the one-phonon scattering probabilities  $P_{1}$ in Fig. \ref{nonlinscfg2}. As expected, the exact linear coupling result slightly deviates from the exact nonlinear coupling results as it does not pick up all the processes that are present when the nonlinear coupling terms are switched on.
Here we also show the one-phonon scattering probability calculated in the DWBA applied to the full nonlinear coupling interaction $V(z,Z)$. For very low incoming energies near the one-phonon excitation threshold $E_{i}=\hbar\omega_{0}=2.7$ meV, i.e. in the one-phonon scattering limit in which linear coupling yields the dominant contribution, the scattering probability calculated in the DWBA with full nonlinear coupling is in a good agreement with the other ones. However, with the increase of $E_{i}$ the deviations from the exact result soon become large and signify the break down of this approximation due to the nonunitary treatment of all the scattering events induced by $V(z,Z)$.    

Very similar trends are observed in the behaviour of two-phonon scattering 
probabilities ($P_{2}$ in Fig. \ref{nonlinscfg2}). Again, 
the EBA results are in a very good agreement with the exact ones and hence 
this formalism proves to be a very reliable approximate method for treating 
multiphonon scattering in the quantum regime. 
This property allows us to use the analytical form of the EBA scattering probabilities\cite{plrep} to demonstrate that the maxima of $P_{1}(E_{i})$ and
$P_{2}(E_{i})$ appear approximately at the values of $E_{i}$ for which
$P_{1}=P_{0}$ and $P_{2}=P_{1}$, respectively, and that 
$P_{1}(E_{i})$ is largely given by  
$P_{1}(E_{i})=n(E_{i})\exp[-n(E_{i})]$ where
$n(E_{i})=-\ln P_{0}(E_{i})$ is the mean number of phonons excited in the scattering event\cite{plrep}.

Qualitatively very similar results are retrieved for Ne atom 
scattering treated in the same model (see Fig. \ref{nonlinscfg3}). Likewise the case of He scattering, all the unitary treatments yield quantitatively similar results which follow the general trends present in Fig. \ref{nonlinscfg2}. The only strong discrepancy between  exact and approximate results appears for the scattering probability calculated in the 
DWBA with nonlinear coupling, that again completely fails to reproduce exact results. 
The centre and lower panels also demonstrate as how due to the larger 
projectile mass the maxima of all the unitary one-phonon and two-
phonon scattering probabilities shift to lower $E_{i}$ than in the case of lighter He atoms. This is readily explained 
by the scaling properties of $n(E_{i})$ which in the quasi-elastic limit of the EBA, here valid for $E_{i}\ga$ 10 meV, is 
very accurately reproduced by the expression 
$n(E_{i})=8u_{0}^{2}m E_{i}/\hbar^{2}$ that is independent of both $\alpha$ and $D$. This causes exponentially 
faster attenuation of the Debye-Waller factor and all phonon excitation 
probabilities for larger projectile masses and fixed incoming energies.

The most striking feature of the results shown in Figs. \ref{nonlinscfg2} and \ref{nonlinscfg3} is a complete failure of the DWBA with full 
nonlinear coupling to describe the two-phonon scattering probabilities. 
In view of this and the earlier established 
excellent agreement between the measured and calculated EBA scattering 
probabilities for several prototype  systems\cite{plrep,XeCu}, the most important conclusion that can be drawn from Figs. \ref{nonlinscfg2} and \ref{nonlinscfg3} is that 
{\em in inelastic low energy atom-surface scattering the main contribution to the two- and higher order phonon scattering probabilities comes from successive one-phonon scattering events and not from the processes of simultaneous multiple  phonon excitations from a single interaction vertex that arise in nonlinear coupling models}. 
This clearly demonstrates that unlike the case of inelastic neutron scattering from crystal lattices, the   
application of DWBA combined with nonlinear coupling to interpret 
the multi-phonon features in thermal energy atom scattering spectra leads to erroneous results and therefore should be avoided.      

In the present case of a one-dimensional scattering model the CC-computations of the scattering probabilities are feasible for both the linear and nonlinear forms of coupling. Therefore, in such cases it is advantageous to immediately resort to the full nonlinear coupling in the CC-calculations.  This ceases to be the case with the treatments of nonlinear coupling by other methods that yield scattering probabilities fulfilling the unitarity requirement. Hence, the finding that the EBA results are in a very good agreement with the exact ones becomes particularly important for the treatment of full three-dimensional scattering problems involving dispersive phonons for which the CC-calculations may not be feasible yet. 
In view of this, the present results affirm the EBA as a powerful tool for reliable interpretation of low energy atom scattering from surface phonons\cite{plrep}.

In summary, we have investigated the effects of different forms of linear and nonlinear atom-phonon couplings on inelastic atom-surface scattering. By comparing the results of several exact numerical and approximate analytical treatments of a simple model of atom scattering from surface vibrations we have demonstrated that in the regime of thermal energy atom scattering the linear coupling interaction treated to all orders gives a dominant contribution to the scattering probabilities.

\begin{figure}
\caption{Diagrams for the diagonal components of the scattering matrix whose imaginary parts yield the total probability for the processes induced by the depicted phonon emission and absorption vertices. Full and wavy lines denote the atom and one-phonon propagators, respectively. Open circles, full circles and full squares denote vertices proportional to $(u_{0}\alpha)$, $(u_{0}\alpha)^{2}$, and $(u_{0}\alpha)^{4}$, respectively. 
(a) One-phonon scattering process induced by the linear coupling term $ZV_{1}(z)$ treated to second order. (b) A two-phonon scattering process encompassing one "static vertex" $V_{2}(z)$ renormalized by a phonon closed loop. This diagram is induced by the quadratic term $Z^{2}V_{2}(z)$ treated to third order. (c) A two-phonon scattering process with renormalized two-phonon vertices that is induced by the quartic coupling $Z^{4}V_{4}(z)$ treated to second order.}
\label{nonlinscfg1}
\end{figure}

\begin{figure}
\caption{Scattering probabilities for elastic (upper panel), one-phonon inelastic (centre panel) and two-phonon inelastic (lower panel) He atom scattering from a vibrating Xe atom on a cold surface, calculated for the three types of interactions within the scattering formalisms discussed in the text and shown as functions of the incoming He atom energy $E_{i}$.  Solid line: CC result for full $V(z,Z)$; dotted line: CC result for $ZV_{1}(z)$; dashed line: CC result for $ZV_{1}(z)+Z^{2}V_{2}(z)$; dashed-dotted line: EBA result; chained squares: DWBA result for full $V(z,Z)$. Note that $P_{0}=1$ and $P_{1}=0$ below the one-phonon excitation threshold at $\hbar\omega_{0}$, and $P_{2}=0$ below the two-phonon excitation threshold at $2 \hbar\omega_{0}$.}
\label{nonlinscfg2}
\end{figure}

\begin{figure}
\caption{Same as in Fig. \protect\ref{nonlinscfg2} but for Ne atom scattering from a Xe atom on a cold Cu surface. Comparison with Fig. \protect\ref{nonlinscfg2} shows how the scattering probabilities scale with the projectile mass when all other parameters are fixed.}
\label{nonlinscfg3}
\end{figure}

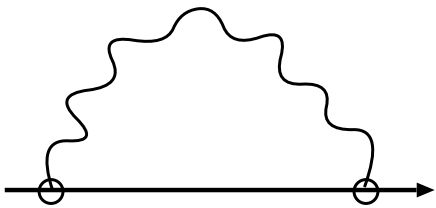
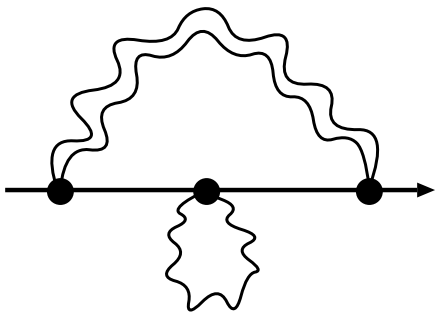
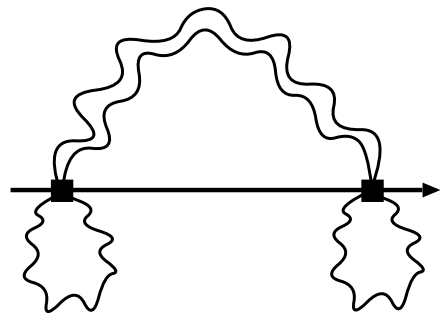
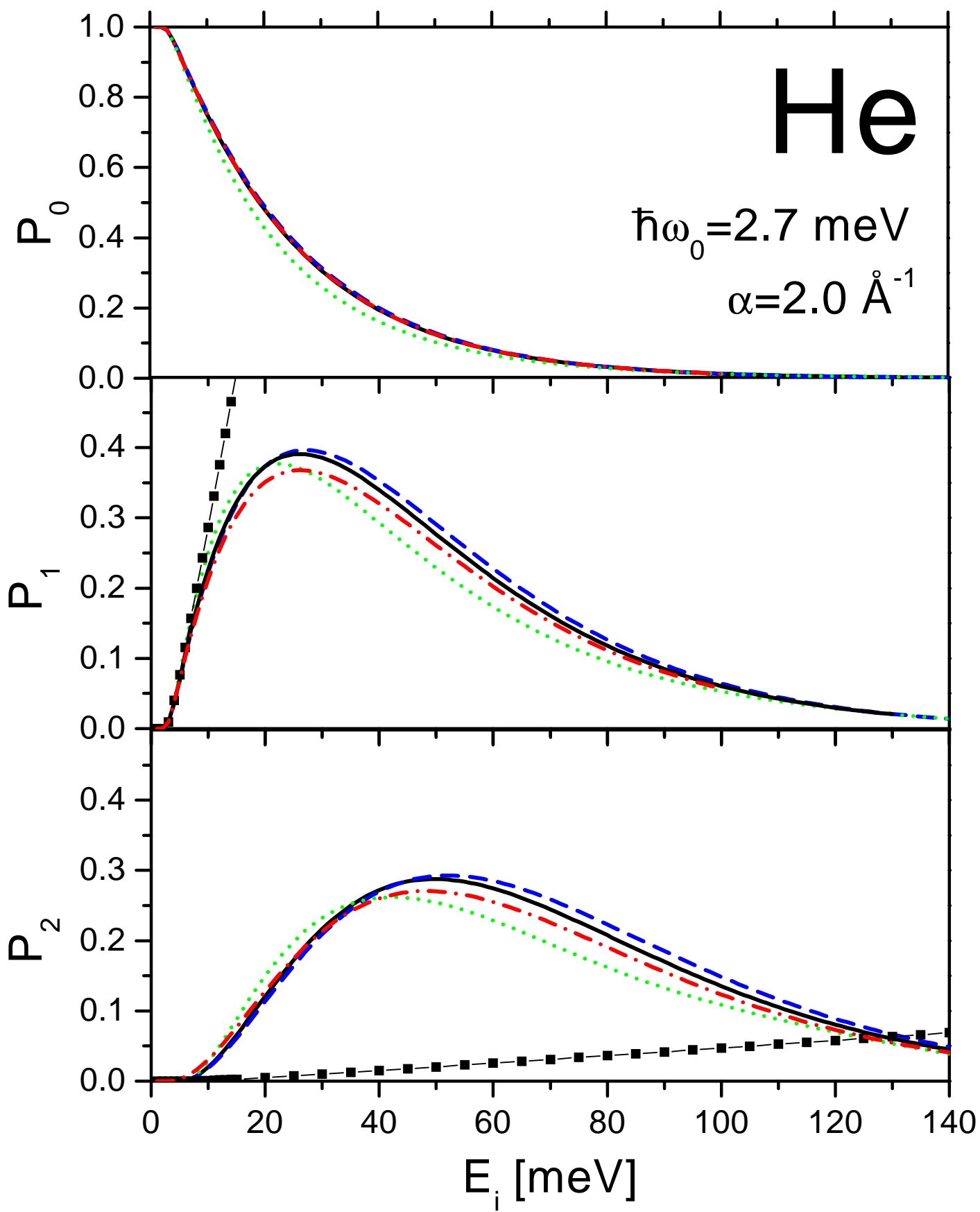
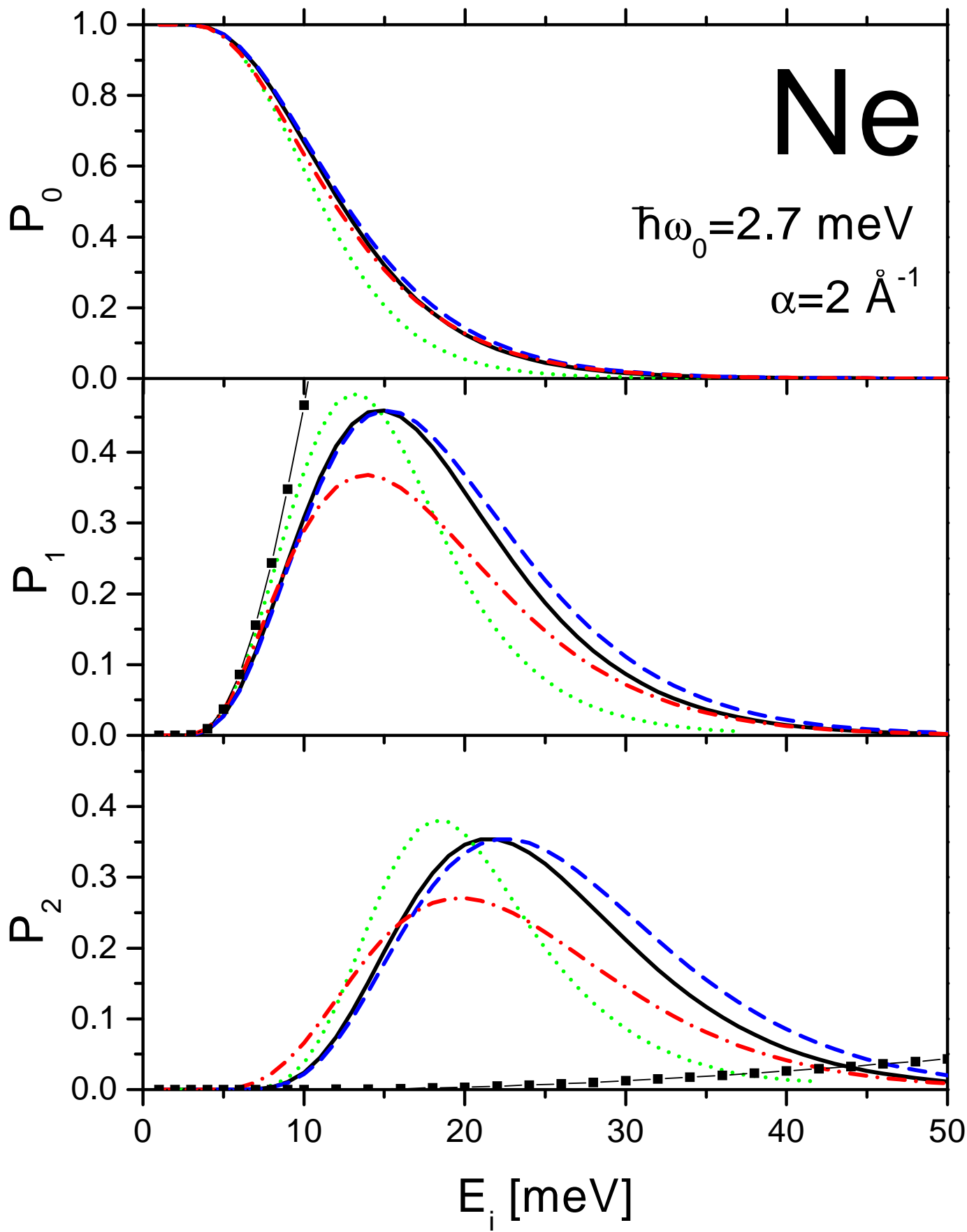
\end{document}